# Disclosing Early Excited State Relaxation Events in Prototypical Linear Carbon Chains

*Piotr Kabaciński,[1] Pietro Marabotti,[2] Daniele Fazzi,[3]\* Vasilis Petropoulos,[1] Andrea Iudica,[1] Patrick Serafini,[2] Giulio Cerullo,[1,4] Carlo S. Casari,[2]\* Margherita Zavelani-Rossi,[2,4]\**

[1] Dipartimento di Fisica, Politecnico di Milano, piazza Leonardo da Vinci 32, 20133 Milano, Italy

[2] Dipartimento di Energia, Politecnico di Milano, via G. Ponzio 34/3, 20133 Milano, Italy

[3] Dipartimento di Chimica "Giacomo Ciamician", Università degli studi di Bologna, via F. Selmi 2, 40126 Bologna, Italy

[4] Istituto di Fotonica e Nanotecnologie IFN-CNR, piazza Leonardo da Vinci 32, 20133 Milano, Italy

*e-mail address: margherita.zavelani@polimi.it; carlo.casari@polimi.it; daniele.fazzi@unibo.it

ABSTRACT

One-dimensional (1D) linear nanostructures comprising *sp*-hybridized carbon atoms, as derivatives of the prototypical allotrope known as carbyne, are predicted to possess outstanding mechanical, thermal, and electronic properties. Despite recent advances in the synthesis, their chemical and physical properties are still poorly understood. Here, we investigate the photophysics of a prototypical polyyne (i.e., 1D chain with alternating single and triple carbon bonds), as the simplest model of finite carbon wire and as a prototype of *sp*-carbon-based chains. We perform transient absorption experiments with high temporal resolution (<30 fs) on monodispersed hydrogen-capped hexayne H−(C≡C)$_6$−H synthesized by laser ablation in liquid. With the support of detailed computational studies based on ground state density functional theory (DFT) and excited state time-dependent (TD)-DFT calculations, we provide a comprehensive description of the excited state relaxation processes at early times following photoexcitation. We show that the internal conversion from a bright high-energy singlet excited state to a low-lying singlet dark state is ultrafast and takes place with a 200-fs time constant, followed by thermalization on the picosecond timescale and decay of the low-energy singlet state with hundreds of picoseconds time constant. We also show that the timescale of these processes does not depend on the end groups capping the *sp*-carbon chain. The understanding of the primary photo-induced events in polyynes is of key importance both for fundamental knowledge and for potential optoelectronic and light-harvesting applications of low dimensional nanostructured carbon-based materials.

INTRODUCTION

Linear nanostructures comprising *sp*-hybridized carbon atoms are one-dimensional (1D) materials (i.e., one-atom-thick chains) featuring peculiar structure-property relationships, as a result of extended π-electron conjugation and strong electron-phonon coupling [1]. The ideal infinite system (i.e., carbyne) represents the lacking carbon allotrope beside graphite and diamond and is predicted to possess outstanding mechanical, thermal and electronic properties [1]-[5]. *sp*-hybridization allows to maximize the π-electron conjugation, leading to two possible structural and electronic configurations: *i*) the equalized bond structure (i.e., the cumulenic form =(C=C)$_n$=, with a sequence of double bonds) as an ideal 1D metal showing a zero band-gap, *ii*) the alternated bond structure (i.e., the polyynic form –(C≡C)$_n$–, with a sequence of alternating single and triple bonds) [6], which is favored by Peierls' instability and features a finite band-gap.

While carbyne has been the focus of many theoretical studies, its experimental realization through chemical or physical processes is still in its infancy. The closest available system to the ideal carbyne is a long linear carbon chain encapsulated in the core of a double-walled carbon nanotube, the so-called confined carbyne [7]. This system has been shown to feature the strongest resonance Raman cross section ever reported, with a very large Huang-Rhys factor indicating a strong electron-phonon coupling [8]-[9]. However, the electronic and optical characteristics of confined carbyne, such as the band-gap and the vibrational frequencies of the Raman-active mode, depend on the type of encapsulating system, making it difficult to assess the intrinsic carbyne properties by looking for the saturation of size-dependent effects [11]-[16]. Size- (chain length) and termination-(end-capping groups) dependent effects are indeed dominating in finite-size *sp*-carbon chains or carbon atomic wires, making such systems attractive for developing materials with tailored properties [17]-[19]. For example, a field-effect transistor with promising field-effect charge mobility has been recently realized using short cumulenic wires (i.e., three C=C bonds) as active material [20][21].

The strong relationship between structural, electronic and vibrational properties in linear carbon chains has been exploited for their characterization mainly through UV-vis absorption and Raman spectroscopy. Recently, resonant Raman in the UV in combination with UV-vis absorption has been used to retrieve the electronic and vibrational structure of the ground and first excited states of short polyynic chains, showing a peculiar size-dependent electron-phonon coupling [22]. This was possible thanks to the neat vibronic effects characterizing the absorption spectrum. Steady-state Raman and UV-vis absorption spectroscopy, however, do not provide any information about the properties of the excited states.

Up to now, only few studies reported on the excited state dynamics of carbynes [12],[23],[24]. Due to the difficulties in synthesizing simple and stable systems and performing high temporal resolution ultrafast spectroscopy in the challenging UV spectral range, only partial information on specific systems could be obtained. Fazzi *et al.* [23] studied dinaphthyl end-capped polyynes revealing, by visible ultrafast transient absorption (TA) spectroscopy, the inter-system crossing (ISC) process which populates a triplet state from an excited singlet state, with 30 ps formation time. Movsisyan *et al.* [24] used UV-NIR-IR TA spectroscopy to study the excited-state dynamics of an hexayne chain, stabilized with aryl-based end groups (i.e., tris(3,5-di-t-butylphenyl)methyl), either free or threaded through a phenanthroline-rotaxane macrocycle. The system showed the formation of a dark $S_1$ excited state via an internal conversion (IC) process from higher-lying singlet states ($S_n$), followed by slow ISC to the triplet state. IC was characterized by a few-ps time constant (~1 ps for the free hexayne, 2-3 ps for the rotaxane-encapsulated chain), and ISC by ~0.4 ns time constant. Very recently [12] a study on oligoynes with different terminal groups and chain lengths (i.e., number of triple bonds $n$ = 4-12) showed, through ultrafast TA spectroscopy, short-lived $S_n$ excited states decaying to $S_1$ with time constants between 1.4 and 8.8 ps (depending on $n$ and on the end group), followed by ISC towards the triplet state on longer time scales (0.1-4.7 ns).

Overall, the investigated systems have bulky sterically hindered end groups (e.g., aryl-based) [25] that are needed to stabilize the *sp*-chain, however possibly perturbing the overall excited state energies and potential energy profiles when compared with the prototypical (ideal) hydrogen-capped system H–(C≡C)$_n$–H. The questions about the intrinsic excited state properties of the *sp*-carbon backbone and the influence of end groups on the relaxation dynamics of the carbon wires are still open. Moreover, despite the expected ultrafast dynamics due to the extended π-electron conjugation, the early events, characterizing the fate of the photoexcited singlet state and the IC process occurring on the 100-fs timescale, have not been addressed yet. In this context, hydrogen-capped polyynic *sp*-carbon chains represent a prototypical case of finite-size carbyne with the simplest possible termination and can be exploited as a model system to investigate the intrinsic photophysical properties of short *sp*-carbon chains.

It has been proposed that excited state dynamics of polyynes bears similarity to that of carotenoids, which are conjugated $sp^2$-carbon chains consisting of alternating single and double bonds. Carotenoids are found in all photosynthetic light-harvesting complexes and play many key roles, including harvesting of the blue-green components of the solar spectrum, photoprotection by quenching singlet oxygen and regulation of the photosynthetic activity through the so-called non-photochemical quenching mechanism [26]. Carotenoids display a peculiar photophysics, whereby the lowest energy excited state $S_1$ has the same symmetry (e.g., $^2A_g$) as the ground state (e.g., $^1A_g$) and is therefore optically dark [27]. Generally, photoexcitation reaches the higher-lying dipole allowed excited state $S_n$ (the $B_{1u}$ state), from which an ultrafast IC to the dark state $S_1$ occurs on the ~100-fs timescale, possibly mediated by intermediate states [27]-[29]. Despite the expected similarity, the complete photoexcitation scenario has not yet been analyzed and understood for *sp*-carbon systems such as polyynes, and in particular for the simple hydrogen-capped (unsubstituted) carbon chain.

Here we perform UV ultrafast TA spectroscopy with <30fs time resolution and broad spectral coverage on the hydrogen-capped polyyne H–(C≡C)$_6$–H, as the simplest possible model of finite *sp*-carbon wire. Our experiments are combined with a detailed computational study based on ground state density functional theory (DFT) and excited state time-dependent (TD)-DFT calculations. We disclose the primary events following the photoexcitation of polyynes, fully characterizing the photophysics of these prototypical carbon atomic wires. We find that photoexcitation reaches a high-lying bright state ($S_n$) and that a very fast IC to a lower-lying dark state ($S_1$) takes place with a ~200-fs time constant. Intraband thermalization processes within the dark state (~2.5 ps time constant) are also observed. Finally, to complete the description of excited state dynamics we also measured the decay of the low-lying singlet state (characterized by a ~500 ps time constant). We found a similar photoexcitation scenario in *sp*-chains with different end groups, such as the mono cyano- H–(C≡C)$_6$–CN and methyl-capped H–(C≡C)$_6$–CH$_3$ species. Our results, enabled by the combination of the synthesis of stable polyynes, a TA setup with high temporal resolution in the UV and quantum-chemical modelling of the TA spectra, provide a comprehensive picture of the primary excited state relaxation events in prototypical carbon atomic wires.

EXPERIMENTAL SECTION AND METHODS

**Synthesis.** Based on the method proposed in [30],[31] , size- and termination-selected polyynes were collected through reversed-phase high-performance liquid chromatography from a polyyne mixture obtained from the ablation of a graphite target in acetonitrile. Employing this method, we obtained three different water/acetonitrile (10/90 v/v) solutions containing, respectively, the hydrogen- (H–(C≡C)$_6$–H), methyl- (H–(C≡C)$_6$–CH$_3$), and cyano-capped H–(C≡C)$_6$–CN polyynes, whose concentrations were estimated from UV-vis absorption spectra (optical path: 1 cm) to be $6.61 \times 10^{-7}$ mol/L (see Figure 1a), $6.38 \times 10^{-8}$ mol/L (see Figure 1b), and $1.42 \times 10^{-7}$ mol/L (see Figure 1c), respectively [32]. For simplicity we will call H–(C≡C)$_6$–H: HC$_{12}$H, H–(C≡C)$_6$–CH$_3$: HC$_{12}$CH$_3$, and H–(C≡C)$_6$–CN: HC$_{12}$CN. The samples have been measured immediately after their collection through chromatography. Previous studies assure their stability in solution for the time required to perform the analysis [31].

**Computational methods.** The ground state structure of HC$_{12}$H was optimized at the DFT level adopting the range-separated functional ωB97X-D3BJ with the triple-split basis set def2-TZVP. The Resolution of Identity approximation, namely J for Coulomb integrals and the COSX numerical integration for Hartree-Fock exchange, was applied as implemented in ORCA v. 5.0.3 [33]. Excited state (singlet) calculations were performed both at the TD-DFT and Tamm-Dancoff (TDA) levels of approximation. Ground to excited state vertical transitions (S$_0$->S$_m$) were computed at the equilibrium ground state geometry i.e., in the Franck-Condon (FC) region. Excited-to-excited state transitions (S$_n$->S$_m$) were evaluated at the TDA level on top of the ground state (S$_0$) geometry. Geometry optimization of the S$_1$ state was performed at the TD-DFT level. S$_1$->S$_m$ transitions were also computed (TDA level) on top of the S$_1$ equilibrium geometry. All calculations were performed considering the $D_{\infty h}$ symmetry point group of H–(C≡C)$_6$–H. The first triplet state (T$_1$) was optimized at the unrestricted (U)DFT level and the T$_1$->T$_m$ transitions were computed both at the TD-DFT and TDA levels. Ground and excited state calculations have been also performed by considering a double-hybrid B2PLYP DFT functional [34], to incorporate the second order Møller-Plesset (MP) corrections to the energies. Data and comparison amongst functionals (ωB97X-D3BJ vs. B2PLYP) and methods (TD-DFT and TDA) are reported in the SI.

For the cases of methyl-(HC$_{12}$CH$_3$) and cyano-(HC$_{12}$CN) capped polyynes, the molecular structures of both singlet and triplet ground state (S$_0$ and T$_1$) were optimized with ωB97X-D3BJ and def2-TZVP basis set. Singlet ground to excited state transitions (S$_0$->S$_n$) as well as singlet and triplet excited state-to-state transitions (namely S$_i$->S$_j$, T$_i$->T$_j$), were computed at the TD-DFT and TDA levels.

**Ultrafast transient absorption spectroscopy.** Ultrafast TA experiments were performed using a home-built setup [35], based on an amplified Ti:Sapphire laser (Libra, Coherent) generating 100-fs pulses at 800 nm central wavelength (1.55 eV) and 1 kHz repetition rate. A fraction of the laser power was used to pump a broadband visible non-collinear optical parametric amplifier (NOPA). The NOPA output pulses, with spectrum spanning 500-600 nm (2.07-2.48 eV), were compressed to ~10-fs duration by chirped dielectric mirrors and successively frequency doubled in a 20-μm-thick Type I β-barium borate crystal, generating broadband UV pump pulses tunable in the range 250-300 nm (4.13-4.96 eV). The UV pulses were compressed with a MgF$_2$ prism pair to nearly transform-limited sub-20-fs duration, as characterized by two-dimensional spectral interferometry [36]. Broadband probe pulses, covering 320-650 nm (1.91-3.87 eV), were obtained through white light continuum generation by focusing the laser fundamental wavelength in a slowly moving 2-mm-thick CaF$_2$ plate. The instrumental response function of the system, depending on the probe wavelength, is estimated to be 25-30 fs. The pump energy was adjusted to be ≈20 nJ (resulting in a fluence of 80 μJ/cm$^2$). For TA measurements on longer timescales, up to 1.3 ns, ~100-fs pump pulses were used, generated by a narrowband NOPA (~10-nm bandwidth) tuneable in the 2.2 – 2.3 eV (540-565 nm) range, frequency doubled in a 200-μm-thick Type I β-barium borate crystal to yield tunable UV pulses. The sample solutions were poured in a 1-mm-thick cuvette. After the sample, the transmitted probe was sent to a spectrometer (SP2150 Acton, Princeton Instruments) and detected using a linear image sensor driven by a custom-built electronic board (Stresing Entwicklungsburo GmbH) working at the full laser repetition rate [37]. For each probe wavelength, the differential absorption (ΔA) was measured as a function of the pump-probe delay.

RESULTS AND DISCUSSION

The prototypical polyyne that we study is the hydrogen-capped HC$_{12}$H. The UV-vis absorption spectrum (see Figure 1a) is characterized by a 0-0 band peaking at 4.54 eV (273 nm) followed by well-resolved vibronic replicas, whose position is strongly dependent on the *sp*-carbon length and termination [1],[31]. The effect of the terminal groups on the 0-0 transition energy is minimal and it is reported in Figures 1b and 1c, showing the UV-vis spectra of mono methyl (HC$_{12}$CH$_3$) and cyano (HC$_{12}$CN) capped polyyne. Both groups favour the extension of the π-electron conjugation, by lowering the 0-0 transition energy to 4.4 eV (for HC$_{12}$CH$_3$) and 4.3 eV (for HC$_{12}$CN). Furthermore, both mono-capped species present weak absorption bands at lower energies than the 0-0 transition, namely at 4.2 eV for HC$_{12}$CH$_3$, and at 4.0 eV and 4.2 eV for HC$_{12}$CN. These transitions can be assigned to non-Condon (e.g., Herzberg-Teller) effects, which activate via vibronic coupling low-lying dark electronic excited states [38]-[40].

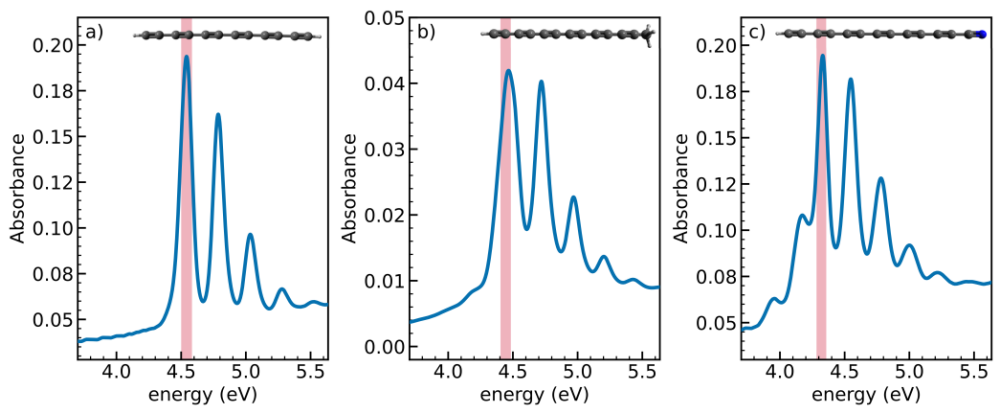

**Figure 1.** UV-vis absorption spectra and chemical structures of $HC_{12}$-R with different end groups (-R) in water/acetonitrile (10/90 v/v) solutions: a) -H, b) -$CH_3$, and c) -CN. Red bands indicate the pump photon energy used in TA experiments.

We investigate the early events of ultrafast excited state relaxation of $HC_{12}H$ using broadband TA spectroscopy in the UV with sub-30-fs temporal resolution. We tune the photon energy of the sub-20-fs pump pulses to match the lowest energy absorption peak (at 4.54 eV, see Figure 1a) and we probe over a broad spectral region (1.9-3.9 eV). Figure 2a shows a map of ΔA as a function of probe photon energy and pump−probe delay, up to 1 ps. We observe two positive bands, corresponding to photo-induced absorption (PA), the first one centered around 3.60 eV (PA1), and the second one around 3.25 eV (PA2) (see also Figure 2b). The PA1 band rises on the sub-50-fs timescale (blue curve in Figure 2c), which is close to our instrumental response function, and decays on the sub-500-fs timescale. The ultrafast decay of the PA1 band corresponds to the rise of the PA2 band, as confirmed by the presence of an isosbestic point around 3.4 eV (see Figure 2b). TA data at time delays up to 30ps are shown in Figure 3. The PA2 band displays a blue shift by

≈0.2 eV from 3.2 to 3.4 eV on the 10-ps timescale (see Figures 3a and 3c), concomitant with a band narrowing and an increase in oscillator strength (see Figure 3c). Similar data were obtained for the other end-capping groups (Figure S1 in SI).

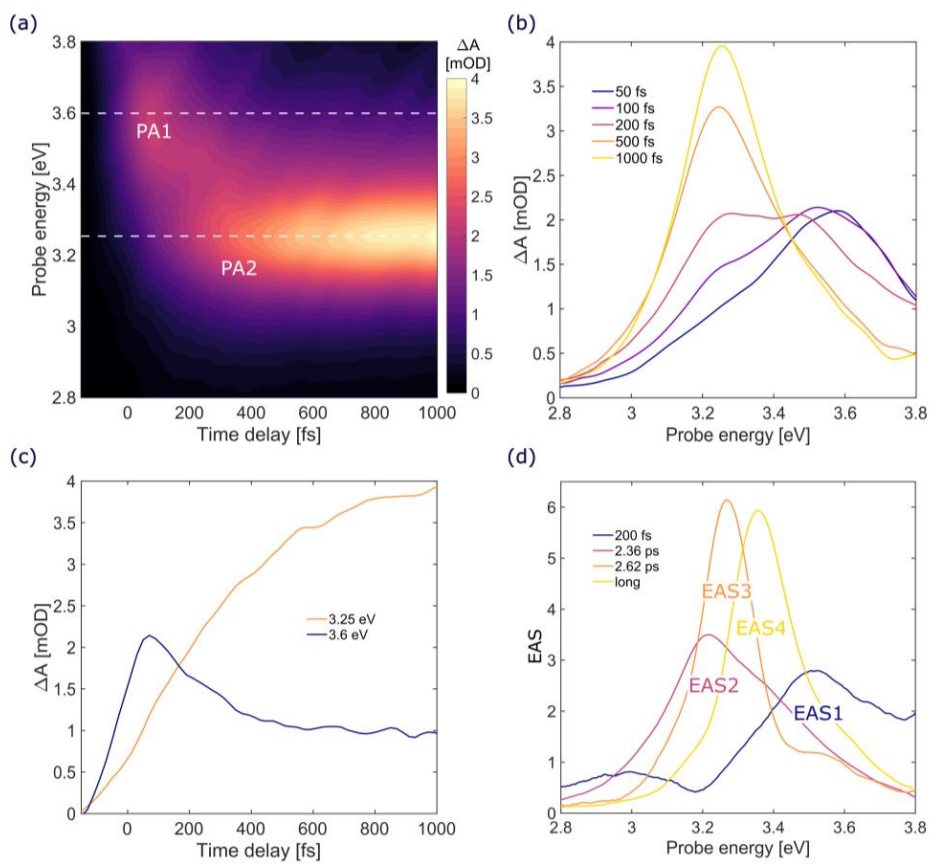

**Figure 2**. (a-c) TA data of HC$_{12}$H in the first picosecond after excitation (pump at 4.54 eV): (a) TA map as a function of delay and probe photon energy; (b) TA spectra at selected pump-probe delays (from 50 fs to 1 ps); (c) TA dynamics at selected probe photon energies, corresponding to the two PA bands (PA1 blue line and PA2 orange line); (d) EAS obtained from the global analysis of the TA data, with the corresponding time constants.

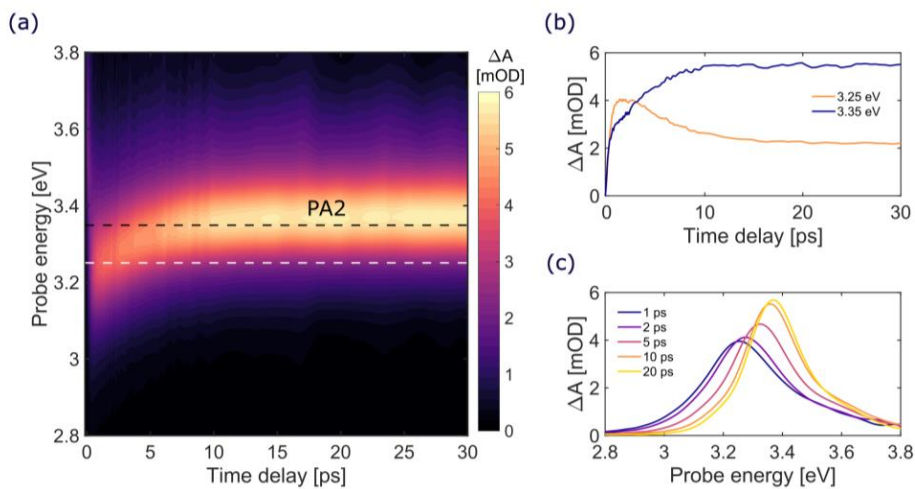

**Figure 3**. TA data of HC$_{12}$H for time delays up to 30ps after excitation (pump at 4.54 eV); (a) TA map as a function of delay and probe photon energy; (b) TA dynamics at selected probe photon energies (3.25 eV orange line, 3.35 eV dark blue line); (c) TA spectra at selected pump-probe delays (from 1 ps to 20 ps).

To obtain further information on the excited state dynamics of the system, we performed a global analysis [41] of TA data using a kinetic model consisting of sequentially interconverting evolution-associated spectra (EAS), with successive mono-exponential decays of increasing time constants, which can be regarded as the lifetimes of each EAS. Results of the global analysis are shown in Figure 2d (and Figure S2 in the SI). We observe a first component (EAS1) corresponding to the PA1 band peaking at ≈ 3.5 eV, which decays with a 200-fs time constant, giving way to EAS2, which corresponds to a "hot" PA2. EAS2 subsequently blue shifts and narrows, evolving into EAS3 and EAS4 with time constants of 2.36 and 2.62 ps respectively. We assign these two EAS rising with similar time constants to the non-exponential dynamics which characterizes the blue-shift and the narrowing associated with the thermalization of the PA2 band. EAS4, which peaks at

≈3.38 eV, accounts for the final "cold" form of PA2 and displays a lifetime which is much longer than our experimental observation window.

Overall, the TA data are consistent with the following scenario. Photoexcitation instantaneously (within the time resolution of our TA apparatus) populates the first bright singlet excited state of $B_{1u}$ symmetry (see TD-DFT calculations below), which gives rise to a PA band (PA1 band in Figure 2a peaking at 3.5 eV) to higher lying singlet states. This PA rapidly decays, giving way to a second, red-shifted PA band (PA2 in Figure 2a), which is assigned to a transition from a dark state ($S_1$) of $A_u$ symmetry to a higher-lying excited state. The matching between the decay of the PA1 band and the growth of the PA2 band is consistent with an IC from the bright ($S_n$) state to a dark ($S_1$) state. In analogy to what wass observed with carotenoids, the IC process is ultrafast, being characterized by a ~200 fs time constant and, according to our experimental results, proceeds without the involvement of intermediate excited states. On longer time scales, the spectral narrowing and blue shift of the PA2 band on the picosecond time scale can be assigned to vibrational relaxation and thermalization of the hot $S_1$ state, which induces a blue-shift and a narrowing of its PA signal. This is again in analogy with the blue shift of the $S_1$ PA band observed in carotenoids and attributed to vibrational cooling [42].

We now compare our experimental data with the quantum-chemical calculations on $HC_{12}H$. Excited states were computed using both TD-DFT and TDA methods, considering for each the range-separated functional ωB97X-D3BJ and the double-hybrid B2PLYP. A detailed comparison amongst methods and functional is reported in the SI. The use of B2PLYP, through the incorporation of static correlation effects via the perturbative MP2 scheme, lowers the energy of the excited states with respect to ωB97X-D3BJ, leading to values that are in good agreement with the experimental data. For such reason, Figure 4a reports the excited energy level diagram of HC$_{12}$H, as computed at the TD-DFT level with B2PYLP functional. The first bright (dipole-allowed) excited state is a high-energy singlet S$_n$ of B$_{1u}$ symmetry, with an energy of 4.59 eV and an oscillator strength f = 5.85. We generally refer here to S$_n$ as a high-lying singlet excited state: for example, it is S$_7$ if computed with TD-DFT and S$_{10}$ if calculated with TDA (see SI). S$_n$ (both at TD-DFT and TDA levels) is characterized by the HOMO->LUMO ($\pi_y$-$\pi^*_y$) and HOMO-1->LUMO+1 ($\pi_x$-$\pi^*_x$) transitions and matches very well the experimental absorption band (Figure 1a) showing the 0-0 vibronic transition centered at 4.54 eV. ωB97X-D3BJ predicts the bright state at a higher energy than B2PLYP, namely at 5.39 eV (f = 6.10), however maintaining the HOMO->LUMO ($\pi_y$-$\pi^*_y$) and HOMO-1->LUMO+1 ($\pi_x$-$\pi^*_x$) character. As well known, ωB97X-D3BJ excited energies are generally overestimated [43]. The first excited state S$_1$ (A$_u$), computed at 2.54 eV at the TD-DFT (B2PLYP) level, is strictly dipole forbidden (f = 0.0) because it involves the HOMO->LUMO+1 ($\pi_y$-$\pi^*_x$) and HOMO-1->LUMO ($\pi_x$-$\pi^*_y$) transitions, which are orthogonal to each other and therefore forbidden by symmetry. The same state is computed at 3.16 eV with ωB97X-D3BJ.

From our analysis, the description of the excited state character of HC$_{12}$H in the FC region is independent of the method (TD-DFT vs. TDA) and functional (B2PLYP vs. ωB97X-D3BJ), all providing very similar results. This point is very relevant for our purposes because it fully justifies the use of TDA (that generally overestimates the excited state energies) for computing the electronic TA spectra by using the code ORCA~~, for which this is the only available computational method~~. For this reason, in the following, we will discuss the computed TA spectra by using the TDA method with the ωB97X-D3BJ functional.

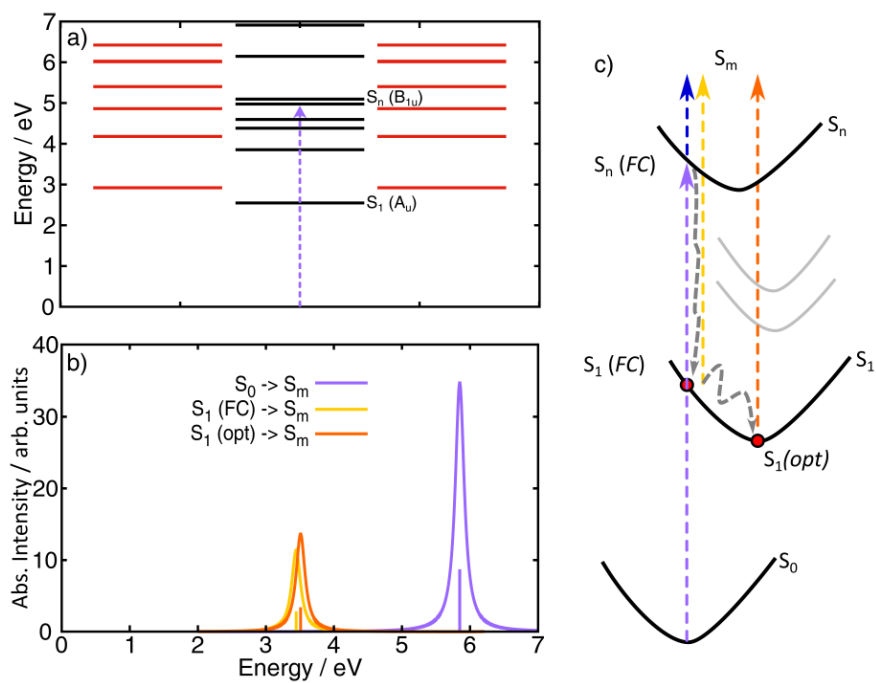

**Figure 4**. (a) Computed TD-DFT (B2PLYP/def2-TZVP) vertical excited state energies of singlet states; red lines show degenerate dark (i.e., symmetry forbidden) excited states, while black lines refer to not degenerate excited states; $S_n$ is the dipole-allowed ($B_{1u}$) excited state (violet arrow represents the transition from the ground state), while $S_1$ is the first (low-lying) dark (symmetry forbidden, $A_u$) state. (b) Computed TDA ($\omega$B97X-D3BJ/def2-TZVP) electronic transitions and spectra for the $S_0 \rightarrow S_m$ transitions (ground-to-high lying excited states) (violet line), the $S_1 \rightarrow S_m$ transitions (transient absorption excited-to-excited states) as evaluated at the FC region (i.e., ground state $S_0$ geometry) (orange line), and the $S_1 \rightarrow S_m$ transitions as calculated at the relaxed (optimized, opt) $S_1$ geometry (dark yellow line); spectra are computed as Lorentzian function centred around the calculated TDA transitions; TDA energies are not scaled. (c) Scheme of the photoinduced relaxation mechanisms involving $S_0$, $S_n$ and $S_1$ states; grey arrows sketch the IC from $S_n$ to $S_1$ (FC region) being the first ultrafast process and the cooling mechanisms within the $S_1$ state. Violet arrow indicates the ground state photoexcitation, blue arrow indicates the PA1 transition, dark yellow and orange arrows the hot and relaxed PA2 transitions, respectively. The experimental energy blue-shift of PA2 (Figure 3) is supported by the computed yellow and orange TA spectra (panel b) and sketched by the corresponding arrows (panel c), respectively.

Figure 4b reports the computed $S_0$->$S_m$ absorption spectrum (violet line) and the TA spectra for two cases, namely i) the $S_1$->$S_m$ transition (dark yellow line), as calculated at the FC region (i.e., at the $S_0$ equilibrium geometry), and ii) the $S_1$->$S_m$ transition (orange line), as computed at the relaxed (optimized) excited state $S_1$ geometry. For the $S_1$->$S_m$ transition at the FC region, an intense absorption band at 3.44 eV is computed. The calculations of the $S_1$->$S_m$ transition at the $S_1$ optimized geometry instead reveal a blue shift (~ 0.07 eV) of the band up to 3.51 eV with an increase of its oscillator strength. The computed $S_1$->$S_m$ PA band and the predicted blue shift upon geometry relaxation within the $S_1$ potential energy surface, well match the observed experimental data (Figure 3c) thus supporting the proposed scenario for the ultrafast IC process.

According to quantum-chemical calculations, $HC_{12}H$ shows different equilibrium geometries and structural relaxations, depending on the populated electronic state (see Figure S3 in the SI). While in the ground state ($S_0$) $HC_{12}H$ shows a clear alternation of single/triple bonds within the *sp*-chain, featuring a bond length alternation parameter (BLA = ($<R_{single}>$ - $<R_{triple}>$)/$n$, averages of single and triple bonds, with $n$ number of bonds) of the order of 0.16 Å, in the $S_1$ state $HC_{12}H$ shows a highly equalized (cumulenic like) structure, resulting in a BLA parameter as low as 0.05 Å in the central part of the *sp*-chain. This is in agreement with previous studies showing a tendency towards equalization upon charge transfer induced by interaction with metal nanoparticles or by the selection of specific end groups [44]-[46]. Furthermore, the calculated TD-DFT (ωB97X-D3BJ) energy of the relaxed $S_1$ state is 2.88 eV, while the energy of the relaxed $T_1$ state is 2.29 eV, which is 0.59 eV below $S_1$. The geometry of $T_1$ is very similar to that of $S_1$, showing as well a BLA parameter of 0.05 Å in the central part of the *sp*-chain. $S_1$->$T_1$ ISC mechanisms might therefore be possible, as already observed for the case of dinaphthyl end-capped polyynes [23], given the lower energy of $T_1$ with respect to $S_1$ and the similarity of the molecular geometries, possibly resulting in effective spin-orbit couplings.

For the methyl- and cyano-capped species ($HC_{12}CH_3$ and $HC_{12}CN$), the computed TD-DFT energies of the bright $S_n$ state shift towards lower values, in good agreement with the experimental data (Figures 1b and 1c). The TD-DFT ($\omega$B97X-D3BJ) $S_n$ vertical energy is 5.39 eV for $HC_{12}H$, 5.33 eV for $HC_{12}CH_3$ and 5.14 eV for $HC_{12}CN$ (unscaled TD-DFT values). Similarly, a red shift of the triplet states occurs for the cyano-capped species. While for $HC_{12}CH_3$ the (relaxed) $T_1$ energy equals that of $HC_{12}H$ (that is 2.29 eV), for $HC_{12}CN$ the $T_1$ minimum red-shifts to 1.88 eV, being the lowest amongst the three species.

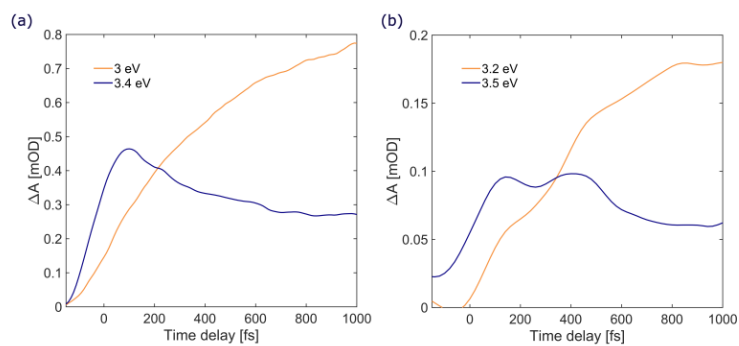

**Figure 5**. TA dynamics at selected probe photon energies for (a) $HC_{12}CH_3$ and (b) $HC_{12}CN$.

It is important to notice that the observed early excited state relaxation events in polyynes are independent of the termination. TA data measured on $HC_{12}CH_3$ (Figure 5a and Figure S1a,b) and $HC_{12}CN$ (Figure 5b and Figure S1c,d), show similar spectroscopic features, i.e. an IC from the bright $S_n$ to the dark $S_1$ state, with comparable time constants. This experimental observation indicates that end-capped groups do not play a role in the early photoinduced decay events; moreover, this further proves that the simplest possible end-capped *sp*-carbon wires, namely hydrogen-capped polyynes $HC_nH$, are the prototypical systems for 1D carbon-based materials and that the knowledge of their photoinduced decay mechanisms unveils the fundamental photophysics of any polyyne-like species.

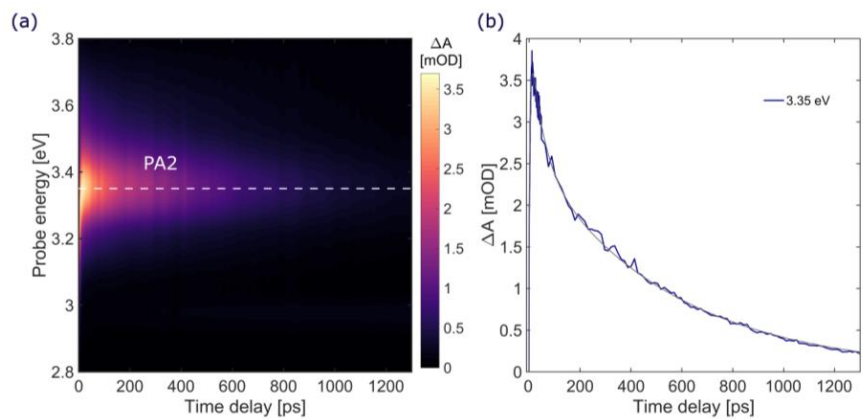

**Figure 6**. TA data on the long timescale, up to 1.3 ns, for $HC_{12}C$. (a) TA map, (b) TA $S_1$ singlet state dynamics at 3.35 eV (blue line) and fitting curve (black line).

Our study focuses on the early excited state relaxation processes in polyynes, nevertheless for completeness we also investigated ~~and we did not investigate deeply~~ the fate of the $S_1$ state. Figure 6a displays the TA map obtained with a narrow-band long (~100 fs) pump pulse, on a long time scale. Here, we can follow the decay of the $S_1$ state, which is characterized by two time constants (~50 ps and 540 ps, Figure 6b). Similar behaviour is observed for all samples (Figure S4). TD-DFT calculations suggest the possibility of an ISC to the triplet state, in particular for the cyano-capped species, which features a $T_1$ energy lower than hydrogen- and methyl-capped chains. We have computed the triplet-to-triplet transitions ($T_1$-> $T_n$) at TD-UDFT level for each species. For $HC_{12}CN$ the first non-negligible $T_1$-> $T_n$ transition is found at 3.72 eV ($T_1$-> $T_{15}$, f = 0.15), while for $HC_{12}H$ and $HC_{12}CH_3$ it is predicted at higher energies (~ 4.00 eV, f = 0.33), out of our experimental probing window. Indeed, for $HC_{12}CN$ we could detect the triplet state PA band, corresponding to a $T_1$-$T_n$ excitation peaking at ≈3.77 eV, in very good agreement with our calculations, and observe that its growth (~500 ps) matches with the decay of the PA of the dark singlet state $S_1$ (Figures S1d, S4c, and S4d), as expected and consistently with polyynes featuring other end-capped groups so far reported (120 ps-1.1 ns [12] and ~ 400 ps [24]).

Further interesting information would be obtained from the detection of impulsively excited coherent oscillations, which could shed light into the vibrational modes coupled to the electronic transitions. However, our temporal resolution enables us to resolve oscillations at frequencies lower than ~ 20 THz (~670 cm$^{-1}$), that are not expected in our systems [9].

CONCLUSIONS

Despite previous attempts to describe the general photophysics of *sp*-carbon chains, such as polyynes (-C≡C-), and access the early events following photoexcitation, experimental difficulties related to the synthesis of such systems and ultrafast spectroscopy allowed only for partial results. Here, by combining the synthesis of prototypical monodispersed hydrogen-capped carbon wires with high temporal resolution ultrafast UV transient absorption spectroscopy and DFT/TD-DFT calculations, we provide a comprehensive description of the primary excited state relaxation processes of polyynes. We prove that the primary photoinduced event is an ultrafast internal conversion from the bright dipole-allowed high-energy singlet state ($S_n$), to a dark low-energy singlet state ($S_1$), and we disclose its ~200 fs time constant. Following internal conversion, a thermalization process in the dark state $S_1$ takes place on a few picoseconds time scale. Eventually, the dark singlet state singlet decays in hundreds of picoseconds with an ISC towards a triplet state. Our results provide a comprehensive description of the primary events of excited state relaxation in prototypical carbon atomic wires, where the electronic and steric perturbations as induced by the end-capping groups are minimized.

ASSOCIATED CONTENT

Supporting information. Supplementary TA maps, EAS data, DFT optimized geometries, TDA and TDDFT excited state energies.

AUTHOR INFORMATION

e-mail address: margherita.zavelani@polimi.it; carlo.casari@polimi.it; daniele.fazzi@unibo.it

ACKNOWLEDGEMENTS

P.M., M.Z.-R., P.S. and C.S.C. acknowledge funding from the European Research Council (ERC) under the European Union's Horizon 2020 research and innovation program ERC-Consolidator Grant (ERC CoG 2016 EspLORE grant agreement no. 724610, website: www.esplore.polimi.it). C.S.C. acknowledges partial funding from the project PNRR - Partenariati estesi - "NEST - Network 4 Energy Sustainable Transition" - PE0000021. D. F. acknowledges partial funding from the National Recovery and Resilience Plan (NRRP), Mission 04 Component 2, Investment 1.5 – NextGenerationEU, Call for tender n. 3277 dated 30/12/2021, Award Number: 0001052 dated 23/06/2022.

References

[1]. Casari, C. S.; Tommasini, M.; Tykwinski, R. R.; Milani, A.; Carbon-atom wires: 1-D systems with tunable properties, *Nanoscale* **2016**, 8, 4414–4435.

[2]. Wang, M.; Lin, S.; Ballistic Thermal Transport in Carbyne and Cumulene with Micron-Scale Spectral Acoustic Phonon Mean Free Path., *Sci. Rep.* **2015**, *5*, 18122.

[3]. Artyukhov, V. I.; Liu, M.; Yakobson, B. I.; Mechanically Induced Metal-Insulator Transition in Carbyne, *Nano Lett.* **2014**, *14* (8), 4224—4229.

[4]. Lang, N. D.; Avouris, P.; Oscillatory Conductance of Carbon-Atom Wires, *Phys. Rev. Lett.* **1998**, *81* (16), 3515–3518.

[5]. Zanolli, Z.; Onida, G.; Charlier, J.-C.; Quantum Spin Transport in Carbon Chains, *ACS Nano* **2010**, *4* (9), 5174–5180.

[6]. Gao Y.; Tykwinski R.R.; Advances in Polyynes to Model Carbyne, *Acc. Chem. Res.* **2022**, 55 (24), 3616–3630

[7]. Shi, L.; Rohringer, P.; Suenaga, K.; Niimi, Y.; Kotakoski, J.; Meyer, J.C.; Peterlik, H.; Wanko, M.; Cahangirov, S.; Rubio, A.; Lapin, Z.J.; Novotny, L.; Ayala, P.; Pichler, T.*;* Confined linear carbon chains as a route to bulk carbyne. *Nature Mater* **2016**, 15 634–639.

[8]. Tschannen, C. D.; Gordeev, G.; Reich, S.; Shi, L.; Pichler, T.; Frimmer, M.; Novotny, L.; Heeg, S.; Raman Scattering Cross Section of Confined Carbyne. *Nano Lett.* **2020**, 20, 6750–6755.

[9]. Milani, A.; Tommasini, M.; Russo, V.; Li Bassi, A.; Lucotti, A.; Cataldo, F.; Casari C.S.; Raman spectroscopy as a tool to investigate the structure and electronic properties of carbon-atom wires, *Beilstein J. Nanotechnol.* **2015**, 6, 480–491

[10]. Martinati, M.; Wenseleers, W.; Shi, L.; Pratik, S.; Rohringer, P.; Cui, W.; Pichler, T.; Coropceanu, V.; Brédas, J.; Cambré, S.; Electronic structure of confined carbyne from joint wavelength-dependent resonant Raman spectroscopy and density functional theory investigations, *Carbon* **2022**, 189, 276-283.

[11]. Gao, Y.; Hou, Y.; Gordillo Gámez, F.; Ferguson, M.J.; Casado, J.; Tykwinski, R.R.; The loss of endgroup effects in long pyridyl-endcapped oligoynes on the way to carbine, *Nat. Chem.* **2020**, 12, 1143–1149.

[12]. Zirzlmeier, J.; Schrettl, S.; Brauer, J.C.; Contal, E.; Vannay, L.; Brémond, É.; Jahnke, E.; Guldi, D.M.; Corminboeuf, C.; Tykwinski, R.R.; Frauenrath, H.; Optical gap and fundamental gap of oligoynes and carbine, *Nat Commun.* **2020**, 11, 4797.

[13]. Chalifoux, W. A.; Tykwinski, R. R.; Synthesis of polyynes to model the sp-carbon allotrope carbyne. *Nature Chem* **2010**, 2, 967–971.

[14]. Heeg, S.; Shi, L.; Pichler, T.; Novotny, L.; Raman resonance profile of an individual confined long linear carbon chain, *Carbon* **2018**, 139, 581-585.

[15]. Heeg, S.; Shi, L.; Poulikakos, L. V.; Pichler, T.; Novotny, L.; Carbon Nanotube Chirality Determines Properties of Encapsulated Linear Carbon Chain, *Nano Lett.* **2018**, 18, 9, 5426-5431.

[16]. Zhang, K.; Zhang, Y.; Shi, L.; A review of linear carbon chains, *Chin. Chem. Lett.* **2020**, 31, 7, 1746-1756.

[17]. Hu, F.; Zeng, C.; Long, R.; Miao, Y.; Wei, L.; Xu, Q.; Min, W.; Supermultiplexed Optical Imaging and Barcoding with Engineered Polyynes. *Nat. Methods* **2018**, *15* (3), 194–200.

[18]. La Torre, A.; Botello-Mendez, A.; Baaziz, W.; Charlier, J. C.; Banhart, F.; Strain-Induced Metal-Semiconductor Transition Observed in Atomic Carbon Chains, *Nat. Commun.* **2015**, *6*, 2–8

[19]. Bryce, M. R.; A review of functional linear carbon chains (oligoynes, polyynes, cumulenes) and their applications as molecular wires in molecular electronics and optoelectronics, *J. Mater. Chem. C* **2021**, 9, 10534-10546.

[20]. Scaccabarozzi, A.D.; Milani, A.; Peggiani, S.; Pecorario, S.; Sun, B.; Tykwinski, R.R, Caironi, M.; Casari, C.S.; A Field-Effect Transistor Based on Cumulenic sp-Carbon Atomic Wires, *J. Phys. Chem. Lett*. **2020**, 11, 1970-1974

[21]. Pecorario, S.; Scaccabarozzi, A.D.; Fazzi, D.; Gutiérrez-Fernández, E.; Vurro, V.; Maserati, L.; Jiang, M.; Losi, T.; Sun, B.; Tykwinski, R.R.; Casari, C.S.; Caironi, M.; Stable and Solution-Processable Cumulenic sp-Carbon Wires: A New Paradigm for Organic Electronics, *Adv. Mater*. **2022** 2110468

[22]. Marabotti, P.; Tommasini, M.; Castiglioni, C.; Serafini, P.; Peggiani, S.; Tortora, M.; Rossi, B.; Li Bassi, A.; Russo, V.; Casari, C.S.; Electron-phonon coupling and vibrational properties of size-selected linear carbon chains by resonance Raman scattering, *Nat. Comm.* **2022** 13:5052

[23]. Fazzi, D.; Scotognella, F.; Milani, A.; Brida, D.; Manzoni, C.; Cinquanta, E.; Devetta, M.; Ravagnan, L.; Milani, P.; Cataldo, F.; Lüer, L.; Wannemacher, R.; Cabanillas-Gonzalez, J.; Negro, M.; Stagira, S.; Vozzi,C.; Ultrafast spectroscopy of linear carbon chains: the case of dinaphthylpolyynes, *Phys. Chem. Chem. Phys.* **2013**, 15, 9384—9391.

[24]. Movsisyan, L.D.; Peeks, M.D.; Greetham. J.M.; Towrie, M.; Thompson, A.L.; Parker, A.W.; Anderson, H.L.; Photophysics of Threaded sp-Carbon Chains: The Polyyne is a Sink for Singlet and Triplet Excitation, *J. Am. Chem. Soc.* **2014**, 136, 17996−18008.

[25]. Chalifoux W.A.; Tykwinski R.R.; Synthesis of polyynes to model the *sp*-carbon allotrope carbine, *Nature Chemistry* **2010**, 2, 967–971

[26]. Frank, H.A.; Cogdell, R.J.; Carotenoids in photosynthesis, *Photochemistry and photobiology* **1996**, 63 (3), 257

[27]. Polívka, T.; Sundström, V.; Ultrafast Dynamics of Carotenoid Excited States−From Solution to Natural and Artificial Systems, *Chem. Rev.* **2004**, 104 (4), 20212072

[28]. Polívka, T.; Sundström, V.; Dark excited states of carotenoids: Consensus and controversy, *Chem. Phys. Lett.* **2009**, 477 (1–3), 1-11

[29]. Cerullo, G.; Polli, D.; Lanzani, G.; De Silvestri, S.; Hashimoto, H.; Cogdell, R.J.; Photosynthetic light harvesting by carotenoids: detection of an intermediate excited state, *Science* **2002**, 298, 2395-2398.

[30]. Tabata, H.; Fujii, M.; Hayashi, S., Doi, T.; Wakabayashi, T.; Raman and surface-enhanced Raman scattering of a series of size-separated polyynes, *Carbon* **2006**, 44, 3168-3176.

[31]. Peggiani, S.; Marabotti, P.; Lotti, R.A.; Facibeni, A.; Serafini, P.; Milani, A.; Russo, V.; Li Bassi, A.; Casari, C.S.; Solvent-dependent termination, size and stability in polyynes synthesized via laser ablation in liquids, *Phys. Chem. Chem. Phys*. **2020**, 22, 26312–26321

[32]. Eastmond, R.; Johnson, T.R.; Walton, D.R.M.; Silylation as a protective method for terminal alkynes in oxidative couplings: A general synthesis of the parent polyynes H(C C)nH (n = 4–10, 12), *Tetrahedron* **1972**, 28 (17), 4601–4616.

[33]. Neese, F.; Software update: The ORCA program system—Version 5.0, *WIREs Comput Mol Sci.* **2022**; *12:e1606*

[34]. Grimme, S.; Semiempirical hybrid density functional with perturbative second-order correlation, *J. Chem. Phys*. **2006**, 124, 034108

[35]. Borrego-Varillas, R.; Ganzer, L.; Cerullo, G.; Manzoni, C.; Ultraviolet Transient Absorption Spectrometer with Sub-20-Fs Time Resolution, *Applied Sciences* **2018**, 8, 989.

[36]. Borrego-Varillas, R.; Oriana, A.; Branchi, F.; De Silvestri, S.; Cerullo, G.; Manzoni, C.; Optimized Ancillae Generation for Ultra-Broadband Two-Dimensional Spectral-Shearing Interferometry, *J. Opt. Soc. Am. B* **2015**, 32, 1851–1855.

[37]. Polli, D.; Luer, L.; Cerullo, G., High-time-resolution pump-probe system with broadband detection for the study of time-domain vibrational dynamics, *Rev. Scientific Instrum.* **2006**, 77, 023103

[38]. Herzberg G., and Teller E.*,* Schwingungsstruktur der Elektronenübergänge bei mehratomigen Molekülen *Z. Phys. Chem. Abt.* **1933** 21, 410.

[39]. Negri F., Orlandi G., Zerbetto F., Quantum-chemical investigation of Franck-Condon and Jahn-Teller activity in the electronic spectra of Buckminsterfullerene, *Chemical Physics Letters* **1988**, 144, 1, 31-37.

[40]. Santoro, F., Lami, A., Improta, R., Bloino, J., Barone, V., Effective method for the computation of optical spectra of large molecules at finite temperature including the Duschinsky and Herzberg–Teller effect: The Qx band of porphyrin as a case study, *J.*

*Chem. Phys.* **2008**, 128, 224311.

[41]. Van Stokkum, I.H.M.; Larsen, D.S.; Van Grondelle, R.; Global and target analysis of time-resolved spectra. *Biochim Biophys Acta* **2004**, 1657, 82–104.

[42]. de Weerd, F.L.; van Stokkum, I.H.M.; van Grondelle, R; Subpicosecond dynamics in the excited state absorption of all-trans-β-Carotene, *Chem. Phys. Lett.* **2002**, 354 (1-2), 38-43

[43]. Jacquemin, D.; Mennucci, B.; Adamo, C.; Excited-state calculations with TD-DFT: from benchmarks to simulations in complex environments, *Phys. Chem. Chem. Phys*. **2011**,13, 16987-16998.

[44]. Milani, A.; Lucotti, A.; Russo, V.; Tommasini, M.; Cataldo, F.; Li Bassi, A.; Casari, C.S.; Charge transfer and vibrational structure of sp-hybridized carbon atomic wires probed by surface enhanced Raman spectroscopy, *Journal of Physical Chemistry C* **2011**, 115 (26), 12836–12843

[45]. Milani, A.; Tommasini, M.; Barbieri, V.; Lucotti A.; Russo, V.; Cataldo, F.; Casari, C.S.; Semiconductor-to-Metal Transition in Carbon-Atom Wires Driven by sp2 Conjugated End Groups*, J. Phys. Chem. C* **2017** 121 (19), 10562–10570

[46]. Milani, A.; Barbieri, V.; Facibeni, A.; Russo, V.; Li Bassi, A.; Lucotti, A.; Tommasini, M.; Tzirakis, M.D.; Diederich, F.; Casari, C.S.; Structure modulated charge transfer in carbon atomic wires *Scientific Reports* **2019**, 9:1648

# Disclosing Early Excited State Relaxation Events in Prototypical Linear Carbon Chains

*Piotr Kabaciński,[1] Pietro Marabotti,[2] Daniele Fazzi,[3]\* Vasilis Petropoulos,[1] Andrea Iudica,[1] Patrick Serafini,[2] Giulio Cerullo,[1,4] Carlo S. Casari,[2]\* Margherita Zavelani-Rossi,[2,4]\**

[1] Dipartimento di Fisica, Politecnico di Milano, p.za Leonardo da Vinci 32, 20133 Milano, Italy
[2] Dipartimento di Energia, Politecnico di Milano, via G. Ponzio 34/3, 20133 Milano, Italy
[3] Dipartimento di Chimica "Giacomo Ciamician", Università degli studi di Bologna, via F. Selmi 2, 40126 Bologna, Italy
[4] Istituto di Fotonica e Nanotecnologie-CNR, piazza Leonardo da Vinci 32, 20133 Milano, Italy
e-mail address: margherita.zavelani@polimi.it; carlo.casari@polimi.it; daniele.fazzi@unibo.it

## SUPPORTING INFORMATION

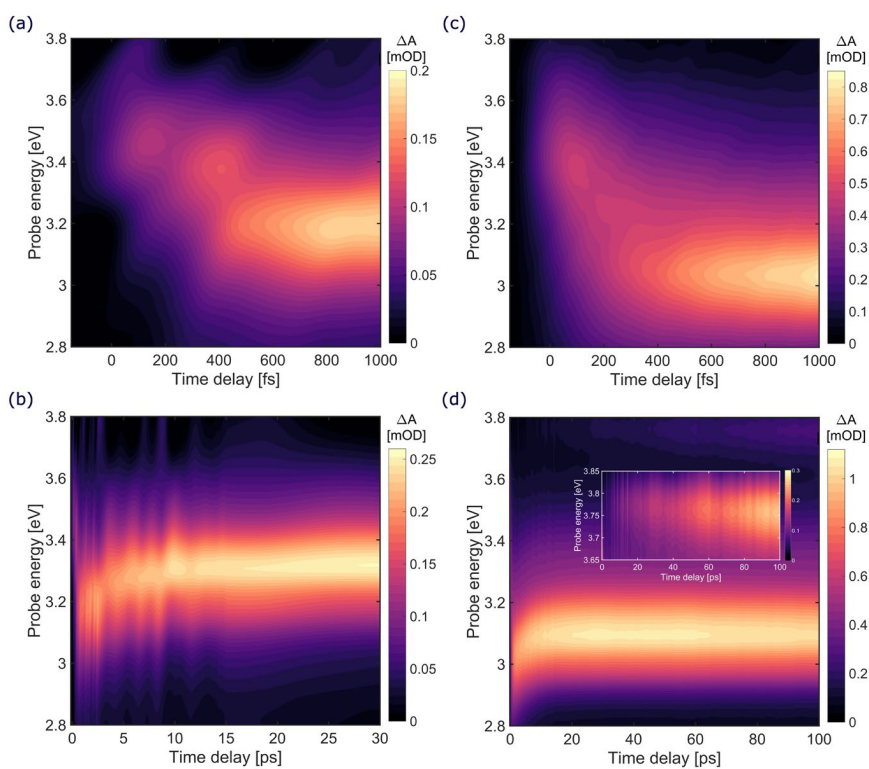

**Figure S1.** Transient differential absorption maps of $HC_{12}$ polyynes with different end groups (methyl and cyano) on different time scales; (a,b) H−(C≡C)$_6$−CH$_3$ pumped at 4.44 eV (279 nm); (c,d) H−(C≡C)$_6$−CN pumped at 4.32 ev (287 nm); inset: zoom out of the 3.65-3.85 eV region, rescaled.

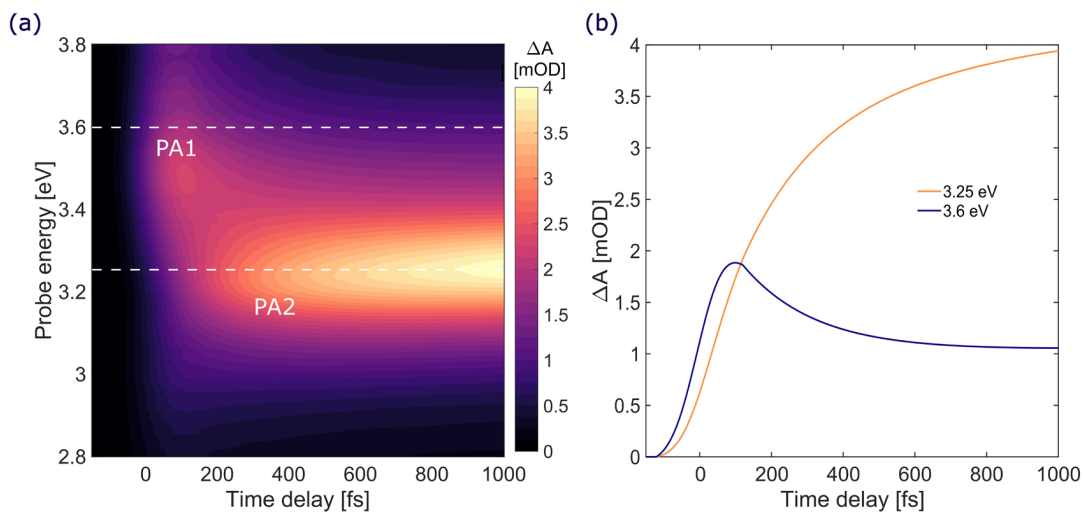

**Figure S2**. (a) EAS obtained from the global analysis of the TA data of HC$_{12}$H; (b) corresponding TA dynamics at selected probe photon energies, corresponding to the two PA bands (PA 1 blue line and PA 2 orange line)

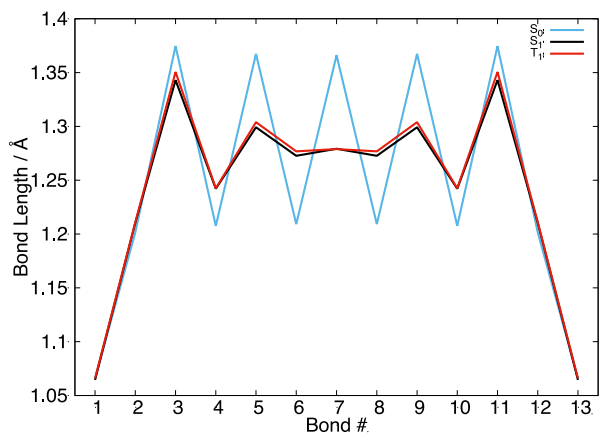

**Figure S3**: Computed equilibrium bond lengths (Å) for HC$_{12}$H in the optimized ground state S0 (light blue line), first (dark) excited state S$_1$ (dark) and triplet state T$_1$ (red). ωB97X-D3BJ data.

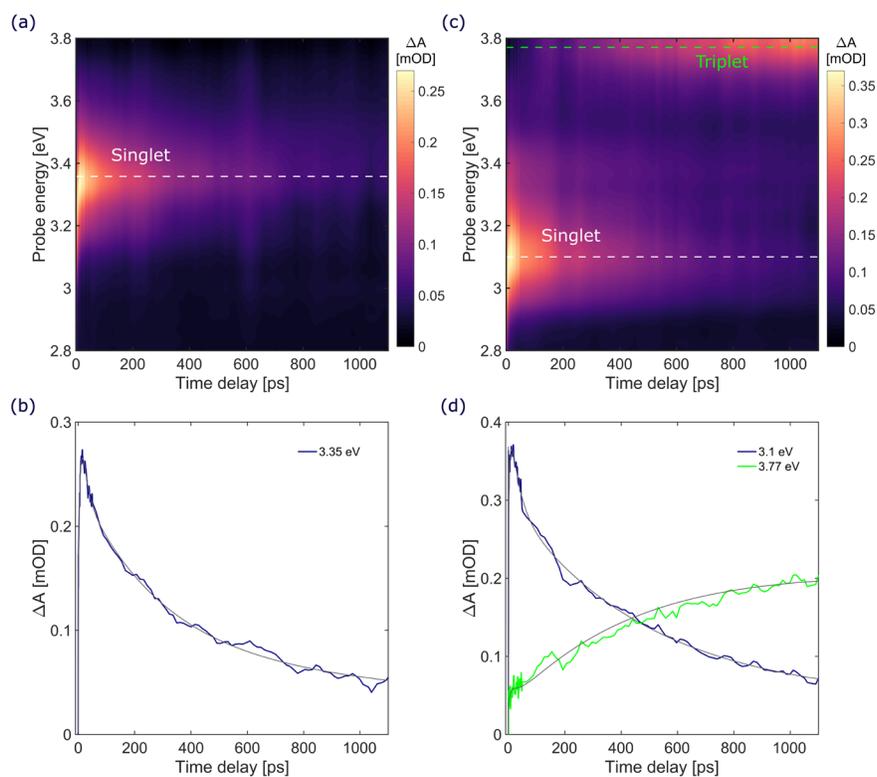

**Figure S4**. TA data on the long time scale, (a,c) maps and (b,d) dynamics for selected wavelengths together with bi-exponential fitting curve obtained by global fitting analysis (black lines) for HC$_{12}$ polyynes with different end groups (methyl and cyano); (a,b) H−(C≡C)$_6$−CH$_3$, time constants ~15 ps, ~290 ps; (c,d) H−(C≡C)$_6$−CN time constants ~30 ps, ~500 ps; here the PA of the triplet state (centered at about 3.77 eV) appears.

**Computational (DFT, TD-DFT and TDA) data**

**Optimized geometries**

Optimized (wB97X-D3BJ def2-TZVP) ground state (S$_0$) geometry for HC$_{12}$H

```
14

  H    0.00000008075489    -0.00000042880016     8.10938967974933
  C    0.00000010571831    -0.00000025841304     7.04344348824528
  C    0.00000009266466    -0.00000008622005     5.84190140146244
  C    0.00000006247950     0.00000008566202     4.46727231805684
  C    0.00000002365473     0.00000022482420     3.25970101252309
  C   -0.00000005310736     0.00000031862900     1.89229803243581
  C   -0.00000012981289     0.00000033558428     0.68308119445412
  C   -0.00000019510128     0.00000026930375    -0.68308103852485
  C   -0.00000020240086     0.00000016683383    -1.89229790643410
  C   -0.00000016324765     0.00000003255490    -3.25970095847324
  C   -0.00000007934853    -0.00000007357661    -4.46727235326350
  C    0.00000003811484    -0.00000015484326    -5.84190148128879
  C    0.00000015553554    -0.00000020417574    -7.04344360021106
  H    0.00000026409611    -0.00000022736312    -8.10938978873136
```

Optimized (B2PLYP/D3 def2-TZVP) ground state (S$_0$) geometry for HC$_{12}$H

```
14

  H   -0.00017207924244     0.00041803745099     8.09680152484801
  C   -0.00011812484375     0.00025037149724     7.03433754015696
  C   -0.00005168089492     0.00007364761763     5.81864303750134
  C    0.00002992449213    -0.00010241542689     4.46722292150165
  C    0.00011768708201    -0.00019886130492     3.23933749583188
  C    0.00013633395143    -0.00028233068579     1.90023072552564
  C    0.00012259981764    -0.00025354994907     0.66805465506586
  C    0.00010489543299    -0.00021831545959    -0.66828260645495
  C    0.00010351774378    -0.00008395681823    -1.90017648682808
  C    0.00004664108122    -0.00007464902566    -3.23925451406323
  C   -0.00004338181496    -0.00015578098689    -4.46714291471396
  C   -0.00009882917516    -0.00006745663819    -5.81857612963456
  C   -0.00009000271994     0.00022886170767    -7.03436453345546
  H   -0.00008750091002     0.00046639802169    -8.09683071528109
```

Optimized (TD-wB97X-D3BJ def2-TZVP) excited state (S$_1$) geometry for HC$_{12}$H

```
14

  H    0.00000046028401    -0.00000259154718     8.07137675529965
```

```
C    0.00000023869049    -0.00000173959360     7.00685729527599
C   -0.00000005266064    -0.00000079610272     5.79658881007522
C   -0.00000034093049     0.00000020592243     4.45354793473270
C   -0.00000045283280     0.00000107329562     3.21145764919539
C   -0.00000035741116     0.00000177176198     1.91230264736089
C   -0.00000015639080     0.00000215454095     0.63957854488649
C    0.00000005865348     0.00000216687461    -0.63957841609106
C    0.00000020930852     0.00000175893411    -1.91230242275419
C    0.00000027165384     0.00000100607638    -3.21145777021034
C    0.00000020740317     0.00000006278390    -4.45354784518386
C    0.00000005382095    -0.00000095750061    -5.79658895642246
C   -0.00000004481622    -0.00000174597510    -7.00685737764865
H   -0.00000009477235    -0.00000236947077    -8.07137684851577
```

Optimized (TD-wB97X-D3BJ def2-TZVP) triplet state ($T_1$) geometry for $HC_{12}H$

```
H    0.00000017481086     0.00000044972401     8.08990840802570
C    0.00000013447646     0.00000024987933     7.02413049709870
C    0.00000008722620     0.00000005312923     5.81312815366413
C    0.00000003216242    -0.00000012014285     4.46266799046614
C   -0.00000001774053    -0.00000022360658     3.22014690508755
C   -0.00000007943104    -0.00000026000466     1.91626229759867
C   -0.00000013577840    -0.00000025907115     0.63943935803695
C   -0.00000020733300    -0.00000024252876    -0.63943930625782
C   -0.00000024576652    -0.00000018523203    -1.91626223818848
C   -0.00000023008352    -0.00000010312039    -3.22014686301095
C   -0.00000010816838     0.00000000337398    -4.46266799611492
C    0.00000004650509     0.00000012304644    -5.81312817665697
C    0.00000020285045     0.00000021563688    -7.02413054535727
H    0.00000034626991     0.00000029891654    -8.08990848439144
```

Optimized (wB97X-D3BJ def2-TZVP) ground state ($S_0$) geometry for $HC_{12}CH_3$

```
H    0.01592364241943    -0.03164115369698     8.91516259220102
C    0.01026673357866    -0.02971290646687     7.84931857462199
C    0.00396429470628    -0.02748672859209     6.64768946107539
C   -0.00294962746468    -0.02466360992110     5.27293075292177
C   -0.00855423465083    -0.02175023737810     4.06525012223177
C   -0.01394200084327    -0.01753820639666     2.69778812884552
C   -0.01753537886202    -0.01272433555582     1.48835976635756
C   -0.01970832035771    -0.00547898697795     0.12226446002656
C   -0.01953841002695     0.00295903659551    -1.08738012150632
C   -0.01637197458588     0.01547736341717    -2.45437905408450
C   -0.01047501673671     0.02961143370833    -3.66284899486258
```

```
  C      0.00063155124838        0.05072651257745       -5.03583727454672
  C      0.01077008725596        0.07346070640377       -6.23950651589081
  C      0.03479235016600        0.10019958844555       -7.69678589421651
  H     -0.95881550666058        0.32329291870180       -8.09187571520119
  H      0.35487801918986       -0.86711407397764       -8.09046064392988
  H      0.72818292282430        0.86547559242720       -8.05260208855686
```

Optimized (UwB97X-D3BJ def2-TZVP) triplet state (T$_1$) geometry for HC$_{12}$CH$_3$

```
17

  H     -8.926176    -0.006608     0.009776
  C     -7.861359    -0.004812     0.007641
  C     -6.649696    -0.003399     0.004996
  C     -5.303435    -0.001714     0.001837
  C     -4.060505    -0.000046    -0.001107
  C     -2.758523     0.000162    -0.003661
  C     -1.483877     0.003835    -0.004323
  C     -0.205653     0.003328    -0.004322
  C      1.070993     0.005205    -0.005224
  C      2.368882     0.006040    -0.005983
  C      3.615592     0.007338    -0.006550
  C      4.956410     0.004930    -0.006557
  C      6.172081     0.000537    -0.009427
  C      7.623592    -0.001691     0.009709
  H      8.016157    -0.693747    -0.738166
  H      7.929575    -0.402925     0.982576
  H      8.073428     0.984998    -0.116355
```

Optimized (wB97X-D3BJ def2-TZVP) ground state (S$_0$) geometry for HC$_{12}$CN

```
15

  H     -0.000001    -0.000001     9.433156
  C     -0.000000    -0.000001     8.367003
  C     -0.000000    -0.000000     7.165622
  C      0.000000     0.000000     5.791331
  C      0.000000     0.000000     4.583844
  C      0.000000     0.000001     3.217029
  C      0.000000     0.000001     2.007843
  C      0.000000     0.000001     0.642685
  C      0.000000     0.000001    -0.566711
  C      0.000000     0.000000    -1.931931
  C      0.000000     0.000000    -3.140514
  C      0.000000     0.000000    -4.507927
  C     -0.000000    -0.000000    -5.713015
  C     -0.000000    -0.000001    -7.092296
  N     -0.000001    -0.000001    -8.244652
```

Optimized (UwB97X-D3BJ def2-TZVP) triplet state (T$_1$) geometry for HC$_{12}$CN

```
15

H        0.000000     0.000000    -9.326275
C        0.000000     0.000000    -8.262042
C        0.000000     0.000000    -7.052671
C        0.000000     0.000000    -5.709223
C        0.000000     0.000000    -4.472288
C        0.000000     0.000000    -3.167922
C        0.000000     0.000000    -1.901532
C        0.000000     0.000000    -0.624852
C        0.000000     0.000000     0.652946
C        0.000000     0.000000     1.935593
C        0.000000     0.000000     3.193262
C        0.000000     0.000000     4.509209
C        0.000000     0.000000     5.736338
C        0.000000     0.000000     7.093003
N        0.000000     0.000000     8.249622
```

## Excited States

Excited state energies (vertical transitions as computed on the optimized wB97X-D3BJ geometry) at the TDA level (first 20 excited states). States S1 and S10 are highlighted for clarity.

```
------------------------------------------------------------------------------
         ABSORPTION SPECTRUM VIA TRANSITION ELECTRIC DIPOLE MOMENTS
------------------------------------------------------------------------------
State   Energy    Wavelength   fosc         T2         TX        TY        TZ
        (cm-1)    (nm)                      (au**2)    (au)      (au)      (au)
------------------------------------------------------------------------------
   1    26514.4    377.2    0.000000000    0.00000    0.00000   -0.00000   0.00000
   2    27136.8    368.5    0.000000000    0.00000   -0.00000   -0.00000  -0.00001
   3    27149.1    368.3    0.000000000    0.00000   -0.00000    0.00000  -0.00001
   4    34500.2    289.9    0.000000000    0.00000   -0.00000    0.00000  -0.00000
   5    35627.7    280.7    0.000000000    0.00000    0.00000    0.00000   0.00001
   6    35641.0    280.6    0.000000000    0.00000   -0.00000    0.00000   0.00001
   7    43419.3    230.3    0.000000000    0.00000    0.00000   -0.00000   0.00000
   8    45057.8    221.9    0.000000000    0.00000   -0.00000   -0.00000   0.00000
   9    45073.0    221.9    0.000000001    0.00000    0.00000   -0.00000  -0.00009
  10    47171.8    212.0    8.724275389   60.88673   -0.00000   -0.00000   7.80299
  11    51807.5    193.0    0.000000000    0.00000    0.00000   -0.00000  -0.00000
  12    53842.9    185.7    0.000000000    0.00000   -0.00000   -0.00000  -0.00001
  13    53861.4    185.7    0.000000000    0.00000    0.00000   -0.00000  -0.00001
  14    54285.0    184.2    0.000000000    0.00000    0.00000    0.00000   0.00001
  15    54289.9    184.2    0.000000000    0.00000   -0.00000    0.00000  -0.00002
  16    54311.0    184.1    0.000000000    0.00000   -0.00000    0.00000   0.00000
  17    55211.1    181.1    0.000000002    0.00000    0.00000    0.00000  -0.00010
  18    58492.8    171.0    0.000000000    0.00000   -0.00000   -0.00000  -0.00000
  19    60759.5    164.6    0.000000000    0.00000    0.00000   -0.00000   0.00000
  20    60781.6    164.5    0.000000000    0.00000    0.00000   -0.00000  -0.00000
```

Excited state energies (vertical transitions as computed on the optimized wB97X-D3BJ geometry) at the TD-DFT level (first 20 excited states). States S1 and S10 are highlighted for clarity.

```
------------------------------------------------------------------------------
         ABSORPTION SPECTRUM VIA TRANSITION ELECTRIC DIPOLE MOMENTS
------------------------------------------------------------------------------
State   Energy    Wavelength   fosc         T2         TX        TY        TZ
        (cm-1)    (nm)                      (au**2)    (au)      (au)      (au)
------------------------------------------------------------------------------
   1    25492.0    392.3    0.000000000    0.00000   -0.00000   -0.00000   0.00000
   2    26373.1    379.2    0.000000000    0.00000   -0.00000    0.00000  -0.00001
   3    26390.3    378.9    0.000000000    0.00000    0.00000    0.00000   0.00001
   4    33886.8    295.1    0.000000000    0.00000   -0.00000   -0.00000   0.00000
   5    35239.8    283.8    0.000000000    0.00000   -0.00000   -0.00000   0.00001
   6    35256.1    283.6    0.000000000    0.00000    0.00000    0.00002  -0.00001
   7    43066.7    232.2    0.000000000    0.00000    0.00001   -0.00000   0.00000
   8    43498.4    229.9    6.104502598   46.20118   -0.00000   -0.00000  -6.79715
   9    44870.7    222.9    0.000000000    0.00000    0.00000   -0.00000  -0.00004
  10    44887.8    222.8    0.000000001    0.00000    0.00000   -0.00000  -0.00010
  11    51593.0    193.8    0.000000000    0.00000   -0.00000   -0.00000   0.00000
  12    53738.3    186.1    0.000000000    0.00000    0.00000   -0.00000   0.00000
```

```
    13    53758.4    186.0    0.000000000    0.00000    0.00000   -0.00000    0.00000
    14    54088.7    184.9    0.000000000    0.00000    0.00000    0.00000    0.00000
    15    54094.6    184.9    0.000000000    0.00000    0.00000   -0.00000    0.00002
    16    54117.1    184.8    0.000000000    0.00000   -0.00001    0.00000    0.00000
    17    54604.9    183.1    0.000000001    0.00000    0.00000    0.00000   -0.00008
    18    58348.0    171.4    0.000000000    0.00000    0.00000    0.00000   -0.00000
    19    60686.7    164.8    0.000000000    0.00000   -0.00000   -0.00000    0.00000
    20    60710.1    164.7    0.000000000    0.00000    0.00000   -0.00000   -0.00000
```

Excited state energies (vertical transitions as computed on the optimized B2PLYP geometry) at the TDA level (first 20 excited states). States S1 and S10 are highlighted for clarity.

```
-----------------------------------------------------------------------------
         ABSORPTION SPECTRUM VIA TRANSITION ELECTRIC DIPOLE MOMENTS
-----------------------------------------------------------------------------
State    Energy    Wavelength   fosc         T2         TX         TY         TZ
         (cm-1)    (nm)                      (au**2)    (au)       (au)       (au)
-----------------------------------------------------------------------------
     1   22742.1   439.7    0.000000000    0.00000    0.00000    0.00000   -0.00000
     2   23340.2   428.4    0.000000000    0.00000    0.00000    0.00000   -0.00000
     3   23344.2   428.4    0.000000000    0.00000    0.00000   -0.00000   -0.00000
     4   32516.8   307.5    0.000000000    0.00000    0.00000    0.00000    0.00000
     5   33606.0   297.6    0.000000000    0.00000   -0.00000    0.00000    0.00000
     6   33609.2   297.5    0.000000000    0.00000   -0.00000    0.00000   -0.00000
     7   42081.1   237.6    0.000000000    0.00000   -0.00000    0.00000   -0.00000
     8   43533.3   229.7    0.000000000    0.00000    0.00000    0.00000    0.00000
     9   43535.4   229.7    0.000000000    0.00000   -0.00000   -0.00000    0.00000
    10   39767.2   251.5    9.057809931   74.98495   -0.00000    0.00000    8.65938
    11   39726.5   251.7    0.000000000    0.00000   -0.00000    0.00000    0.00000
    12   39728.1   251.7    0.000000000    0.00000    0.00000   -0.00000   -0.00000
    13   40033.4   249.8    0.000000000    0.00000   -0.00000    0.00000    0.00000
    14   36068.5   277.3    0.000000000    0.00000   -0.00000    0.00000   -0.00000
    15   50270.0   198.9    0.000000000    0.00000   -0.00000    0.00000    0.00000
    16   51828.0   192.9    0.000000000    0.00000   -0.00000    0.00000    0.00000
    17   51829.2   192.9    0.000000000    0.00000    0.00000   -0.00000    0.00000
    18   55945.1   178.7    0.000000000    0.00000    0.00000   -0.00000    0.00000
    19   49127.6   203.6    0.000000000    0.00000    0.00000   -0.00000    0.00000
    20   48833.2   204.8    0.000000000    0.00000   -0.00000   -0.00000   -0.00001
```

Excited state energies (vertical transitions as computed on the optimized B2PLYP geometry) at the TD-DFT level (first 20 excited states). States S1 and S10 are highlighted for clarity.

```
-----------------------------------------------------------------------------
         ABSORPTION SPECTRUM VIA TRANSITION ELECTRIC DIPOLE MOMENTS
-----------------------------------------------------------------------------
State    Energy    Wavelength   fosc         T2         TX         TY         TZ
         (cm-1)    (nm)                      (au**2)    (au)       (au)       (au)
-----------------------------------------------------------------------------
     1   20551.4   486.6    0.000000000    0.00000   -0.00000    0.00000   -0.00000
     2   23544.3   424.7    0.000000000    0.00000    0.00000    0.00000    0.00000
     3   23556.6   424.5    0.000000000    0.00000   -0.00000    0.00000   -0.00000
     4   31070.7   321.8    0.000000000    0.00000   -0.00000    0.00000   -0.00000
     5   33703.0   296.7    0.000000000    0.00000   -0.00000   -0.00000   -0.00000
```

|    |         |       |             |          |          |          |          |
|----|---------|-------|-------------|----------|----------|----------|----------|
| 6  | 33714.4 | 296.6 | 0.000000000 |  0.00000 |  0.00000 |  0.00000 | -0.00001 |
| **7**  | **37066.8** | **269.8** | **5.850147210** | **51.95854** |  **0.00000** |  **0.00000** | **-7.20823** |
| 8  | 41091.7 | 243.4 | 0.000000000 |  0.00000 | -0.00000 | -0.00000 |  0.00000 |
| 9  | 43553.7 | 229.6 | 0.000000000 |  0.00000 | -0.00000 | -0.00000 |  0.00001 |
| 10 | 43564.1 | 229.5 | 0.000000000 |  0.00000 |  0.00000 | -0.00000 | -0.00004 |
| 11 | 39193.2 | 255.1 | 0.000000000 |  0.00000 | -0.00000 | -0.00000 |  0.00000 |
| 12 | 39194.8 | 255.1 | 0.000000000 |  0.00000 |  0.00000 | -0.00000 |  0.00001 |
| 13 | 40117.9 | 249.3 | 0.000000000 |  0.00000 |  0.00000 | -0.00000 | -0.00000 |
| 14 | 35334.2 | 283.0 | 0.000000001 |  0.00000 |  0.00000 |  0.00000 | -0.00007 |
| 15 | 49565.2 | 201.8 | 0.000000000 |  0.00000 |  0.00000 |  0.00000 | -0.00000 |
| 16 | 51787.4 | 193.1 | 0.000000000 |  0.00000 | -0.00000 |  0.00000 | -0.00000 |
| 17 | 51798.4 | 193.1 | 0.000000000 |  0.00000 |  0.00000 |  0.00000 |  0.00000 |
| 18 | 55751.9 | 179.4 | 0.000000000 |  0.00000 | -0.00003 |  0.00001 |  0.00000 |
| 19 | 48494.3 | 206.2 | 0.000000000 |  0.00000 |  0.00000 | -0.00000 | -0.00000 |
| 20 | 48583.1 | 205.8 | 0.000000000 |  0.00000 |  0.00000 | -0.00000 |  0.00000 |